\documentclass{article}

\usepackage[preprint]{neurips_2024}

\usepackage[utf8]{inputenc} 
\usepackage[T1]{fontenc}    
\usepackage{hyperref}       
\hypersetup{
colorlinks=true,
citecolor=blue,
linkcolor=blue,
urlcolor=blue,
}
\usepackage{url}            
\usepackage{booktabs}       
\usepackage{amsfonts}       
\usepackage{nicefrac}       
\usepackage{microtype}      
\usepackage{xcolor}         
\usepackage{amsmath}
\usepackage{float}
\usepackage{comment}
\usepackage{pdfpages}

\definecolor{pal1}{HTML}{FFD700}
\definecolor{pal2}{HTML}{FFB14E}
\definecolor{pal3}{HTML}{FA8775}
\definecolor{pal4}{HTML}{EA5F94}
\definecolor{pal5}{HTML}{CD34B5}
\definecolor{pal6}{HTML}{9D02D7}
\definecolor{pal7}{HTML}{0024FF}

\usepackage[noabbrev,nameinlink]{cleveref}

\usepackage[noblocks]{authblk}

\usepackage[backend=biber, doi=true, style=chem-angew]{biblatex}
\AtEveryBibitem{
    \clearfield{urlyear}
    \clearfield{urlmonth}
}

\usepackage{graphicx}
\usepackage[font=small,labelfont=bf]{caption}
\usepackage[version=4]{mhchem}

\newcommand{\email}[1]{\texttt{#1}}

\title{Swarm Intelligence for\\ Chemical Reaction Optimisation}

\begin{document}

\author[1,2\dag]{\textbf{R\'{e}mi Schlama}}
\author[1,2*\dag]{\textbf{Joshua W. Sin}}
\author[3]{\textbf{Ryan P. Burwood}}
\author[1]{\textbf{Kurt P\"{u}ntener}}
\author[1]{\textbf{Raphael Bigler}}
\author[2,4*]{\mbox{\textbf{Philippe Schwaller}}}

\affil[1]{Process Chemistry \& Catalysis, Synthetic Molecules Technical Development, \newline
F. Hoffmann-La Roche AG, Basel, Switzerland}
\affil[2]{Laboratory of Artificial Chemical Intelligence (LIAC), EPFL, Lausanne, Switzerland}
\affil[3]{Solid State Sciences, Synthetic Molecules Technical Development, \newline
F. Hoffmann-La Roche AG, Basel, Switzerland}
\affil[4]{National Centre of Competence in Research (NCCR) Catalysis, EPFL, Lausanne, Switzerland}

\maketitle

\vspace{-3.5em}
\begin{center}
\dag These authors contributed equally to this work.\\
*Corresponding authors. Email:
\email{wing\_pong.sin@roche.com}; \email{philippe.schwaller@epfl.ch}
\end{center}
\begin{abstract}
Chemical reaction optimisation is essential for synthetic chemistry and pharmaceutical development, demanding the extensive exploration of many reaction parameters to achieve efficient and sustainable processes. We report $\alpha$-PSO, a novel nature-inspired metaheuristic algorithm that augments canonical particle swarm optimisation (PSO) with machine learning (ML) for parallel reaction optimisation. Unlike black-box ML approaches that obscure decision-making processes, $\alpha$-PSO uses mechanistically clear optimisation strategies through simple, physically intuitive swarm dynamics directly connected to experimental observables, enabling practitioners to understand the components driving each optimisation decision. We establish a theoretical framework for reaction landscape analysis using local Lipschitz constants to quantify reaction space ``roughness'', distinguishing between smoothly varying landscapes with predictable surfaces and rough landscapes with many reactivity cliffs. This analysis guides adaptive $\alpha$-PSO parameter selection, optimising performance for different reaction topologies. Systematic evaluation of $\alpha$-PSO across pharmaceutically relevant reaction benchmarks demonstrates competitive performance with state-of-the-art Bayesian optimisation methods, while two prospective high-throughput experimentation (HTE) campaigns showed that $\alpha$-PSO identified optimal reaction conditions more rapidly than Bayesian optimisation. $\alpha$-PSO combines the predictive capability of advanced black-box ML methods with interpretable metaheuristic procedures, offering chemists an effective framework for parallel reaction optimisation that maintains methodological clarity while achieving highly performant experimental outcomes. Alongside our open-source $\alpha$-PSO implementation, we release $989$ new high-quality Pd-catalysed Buchwald-Hartwig and Suzuki reactions.

\end{abstract}

\section{Introduction}\label{sec:introduction}

Chemical reaction condition optimisation is necessary in chemical synthesis, requiring the resource-intensive exploration of many reaction variables, particularly when optimisation of multiple reaction outcomes is desired~\cite{zhangProcessChemistryScience2006,taylorBriefIntroductionChemical2023}. The increasing demands of the pharmaceutical industry for sustainable and efficient chemical processes towards active pharmaceutical ingredients (APIs) motivate the need for more effective optimisation approaches~\cite{NewRecipeBiopharmaceutical,paulHowImproveRD2010}. Advances in machine learning (ML) have led to the popularisation of Bayesian optimisation as the premier algorithmic method, having been successfully applied to reaction optimisation in multiple experimental case studies and outperforming traditional reaction optimisation methods such as one-factor-at-a-time and experimentalist guided approaches~\cite{shieldsBayesianReactionOptimization2021, torresMultiObjectiveActiveLearning2022,daltonUtopiaPointBayesian2024,slatteryAutomatedSelfoptimizationIntensification2024,christensenDatascienceDrivenAutonomous2021,guoBayesianOptimizationChemical2023,rankovicGOLLuMGaussianProcess2025}. Modern robotic high-throughput experimentation (HTE) platforms have also enabled highly parallel approaches to reaction screening at miniaturised scales, facilitating the simultaneous exploration of many reaction conditions~\cite{gesmundoMiniaturizationPopularReactions2023, buitragosantanillaNanomolescaleHighthroughputChemistry2015,cernakMicroscaleHighThroughputExperimentation2017}. When combined with Bayesian optimisation, HTE enables efficient data-driven search strategies with highly parallel reaction screening capabilities. The effectiveness of algorithm-guided parallel experimentation is exemplified by recent API synthesis optimisation studies achieving substantially compressed cycle times compared to traditional approaches~\cite{sinHighlyParallelOptimisation2025}.

Despite their demonstrated effectiveness, sophisticated ML models such as Bayesian optimisation approaches operate as black-box methods where the underlying decision-making process remains opaque to practitioners~\cite{adachiLoopingHumanCollaborative2024,yoshikawaGaussianProcessRegression2023}. This opacity presents challenges for users seeking to understand why specific reaction conditions are suggested, how model parameters influence recommendations, or diagnose suboptimal performance. While \textit{post hoc} explanation methods attempt to interpret model decisions~\cite{chauExplainingUncertainStochastic2023}, these explanations lack complete fidelity to the true decision-making mechanism and can mislead practitioners~\cite{rudinStopExplainingBlack2019, jacoviFaithfullyInterpretableNLP2020,madsenInterpretabilityNeedsNew2024,marxAreYouSure2023}. These limitations collectively restrict the ability of human practitioners to trust, validate, and align algorithmic suggestions with their scientific expertise and goals~\cite{bouthillierSurveyMachinelearningExperimental2020}, motivating the exploration of more transparent and effective methods. The success of nature-inspired metaheuristic algorithms in chemical applications, from genetic algorithms in catalyst discovery to molecule generation, highlights the potential for such optimisation frameworks~\cite{trippGeneticAlgorithmsAre2023,duMachineLearningaidedGenerative2024,schmidCatalysingOrganocatalysisTrends2024,gaoSampleEfficiencyMatters2022,jensenGraphbasedGeneticAlgorithm2019}. Among metaheuristic approaches, Particle Swarm Optimisation (PSO)~\cite{kennedyParticleSwarmOptimization1995, blackwellImpactCommunicationTopology2019, kohParallelAsynchronousParticle2006, ruedaespinosaNovelComputationalChemistry2023} presents a natural fit for parallel reaction optimisation. PSO's swarm-based architecture mirrors HTE optimisation workflows, where the iterative selection of experimental batches is guided by simple rules. PSO achieves this by replicating swarm intelligence, an emergent property of biological systems where simple local rules, mimicking the foraging behaviours of bees, ants, and fish, lead to effective batch optimisation~\cite{balamanModelingOptimizationApproaches2019,chengPathPlanningObstacle2021}. Like chemists systematically exploring reaction conditions, PSO uses particles to navigate search spaces without functional assumptions, using physically intuitive swarm dynamics where individual particle experience and collective knowledge drive optimisation strategies directly tied to experimental outcomes.

Herein, we introduce a novel swarm intelligence framework, $\alpha$-PSO, for batched multi-objective chemical reaction optimisation that combines PSO and ML, offering both an interpretable and highly performant approach (\Cref{fig:figure_1}). $\alpha$-PSO reconceptualises reaction conditions as intelligent particles collectively navigating the reaction condition search space with physics-based intuition, using simple swarm movement rules with explicitly specified parameters, enhanced by ML acquisition function guidance. Through comprehensive benchmarking studies, we demonstrate $\alpha$-PSO's effectiveness against state-of-the-art Bayesian optimisation and canonical PSO approaches across pharmaceutically relevant multi-objective \textit{in silico} reaction datasets. Further, in prospective experimental HTE optimisation campaigns, $\alpha$-PSO identified optimal reaction conditions more rapidly than Bayesian optimisation, reaching 94 area percent (AP) yield and selectivity within two iterations for a challenging heterocyclic Suzuki reaction and demonstrated statistically significant superior performance over Bayesian optimisation in a Pd-catalysed sulfonamide coupling. To facilitate practical implementation, we exhaustively scan $\alpha$-PSO parameters, providing detailed insights for tuning swarm behaviour across reaction landscapes, enabling chemists to effectively adapt the method to their specific optimisation challenges. $\alpha$-PSO is made available as an open-source repository alongside the complete HTE datasets (989 reactions) from both prospective experimental campaigns in the Simple User-Friendly Reaction Format (SURF)~\cite{nippaSimpleUserFriendlyReaction2024}.

\section{Machine learning-guided particle swarm optimisation}\label{sec:ml-guided-pso}

Canonical PSO uses a swarm of particles to explore a search space in coordinated groups and converge to an optimal solution (see Supplementary Information Section 2.2). Each particle maintains a working memory of its best individually explored position and the swarm's global best position explored thus far. Guided by this information, particles are moved at each iteration, iteratively updating their memory with improved positions, leading the collective swarm towards optimal solutions within the search space~\cite{kennedyParticleSwarmOptimization1995, kohParallelAsynchronousParticle2006, blackwellImpactCommunicationTopology2019}. This memory-guided exploration parallels how chemists accumulate an understanding of a reaction landscape through observed experimental outcomes. While PSO demonstrates effectiveness across various scientific domains~\cite{ruedaespinosaNovelComputationalChemistry2023}, recent developments have pushed toward increasingly advanced ML models for reaction optimisation, such as reinforcement learning and Bayesian optimisation~\cite{wangIdentifyingGeneralReaction2024, shieldsBayesianReactionOptimization2021, torresMultiObjectiveActiveLearning2022, daltonUtopiaPointBayesian2024}. We hypothesised that integrating such ML models within metaheuristic PSO architectures could create a synergistic framework, combining ML's predictive capability with intuitive procedures directly linked to experimental observables. 

\begin{figure}[H]
    \centering
    \includegraphics[width=1\linewidth]{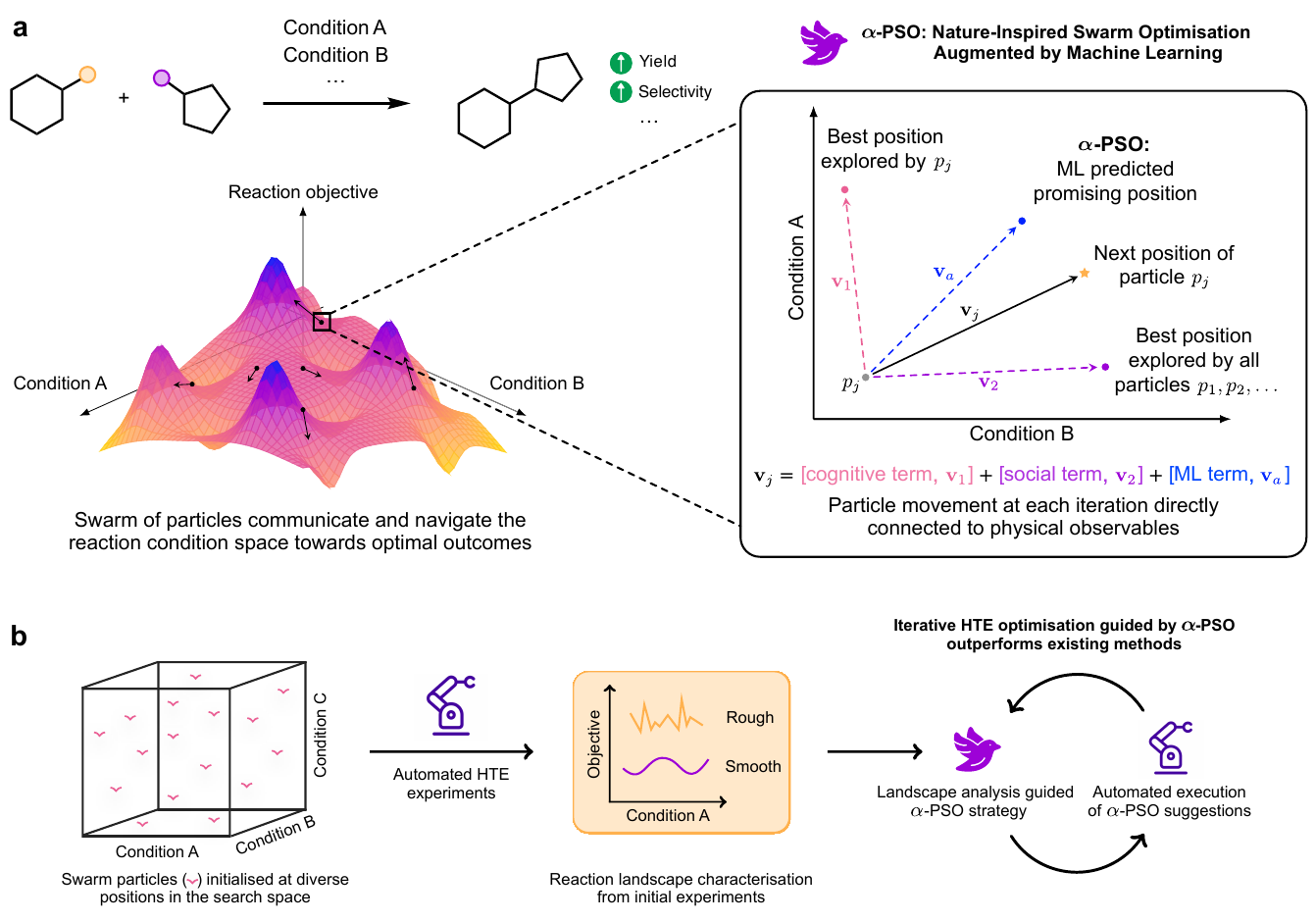}
    \caption{\textbf{$\alpha$-PSO combines nature-inspired swarm intelligence with machine learning (ML) for performant chemical reaction optimisation.} \textbf{a,} Theoretical framework showing how $\alpha$-PSO optimises chemical reactions. Swarm particles navigate the reaction condition landscape spanning arbitrary reaction conditions A and B, towards optimising user-defined reaction objectives (e.g., yield and selectivity). The inset illustrates $\alpha$-PSO's ML-augmented particle movement mechanism, where the trajectory of each particle is determined by its velocity vector. The velocity ($v_j$) of a given particle ($p_j$) is influenced by three components: a cognitive term ($v_1$) directing towards the position of the personal best reaction outcomes explored by the particle ($p_j$), a social term ($v_2$) pointing to the global best position explored by all particles in the swarm, and an ML acquisition function term ($v_a$) guiding particles towards promising positions predicted by ML. The magnitudes of these three terms are weighted by the cognitive coefficient ($c_1$), the social coefficient ($c_2$), and the ML coefficient ($c_\text{a}$), respectively (see Supplementary Information Section 2.3). For clarity in the main text, we refer to these parameters as $c_\text{local}$, $c_\text{social}$, and $c_\text{ml}$, respectively. The physically intuitive movement protocols of $\alpha$-PSO directly connect to explicitly specified regions in the search space, enabling practitioners to easily understand and control optimisation behaviour. \textbf{b,} Experimental application of our iterative $\alpha$-PSO workflow in high-throughput experimentation (HTE) reaction optimisation. Initial quasi-random Sobol sampling populates the reaction condition space with diversely spread particle swarm positions, with automated HTE experiments generating reaction data. Reaction landscape characterisation from these initial experiments classifies the reaction space as either rough (frequent reactivity cliffs) or smooth (more predictable response surfaces), guiding appropriate $\alpha$-PSO parameter configuration. The resulting landscape-guided $\alpha$-PSO strategy iteratively suggests HTE screening plates, optimising the chemical reaction. Applied to two prospective reaction case studies, $\alpha$-PSO outperforms existing state-of-the-art Bayesian optimisation methods in reaction optimisation.}
    \label{fig:figure_1}
\end{figure}

We propose $\alpha$-PSO, a batched optimisation framework where each individual experiment is modelled as an abstract particle that navigates the reaction search space following physics-based swarm dynamics. New batches of reaction condition suggestions are obtained from the iterative, collective movement of the particle swarm. While canonical PSO relies solely on the memory of personal and global best positions~\cite{kennedyParticleSwarmOptimization1995, kohParallelAsynchronousParticle2006, blackwellImpactCommunicationTopology2019}, $\alpha$-PSO augments position update rules with ML acquisition function guidance (\Cref{fig:figure_1}a). This also enables ML predictions to guide strategic particle reinitialisation from stagnant local optima to more promising regions of the reaction space, further extending $\alpha$-PSO's search efficacy (see Supplementary Information Section 2.3). Rather than relying entirely on learned probabilistic abstractions to manage exploration and exploitation, $\alpha$-PSO's decision-making process emerges from direct connections to experimentally observed results. Individual particle experience ($p_\text{best}$, weighted by the cognitive parameter $c_1$) captures local optimisation history while collective knowledge ($g_\text{best}$, weighted by the social parameter $c_2$) represents global search progress, complemented by an ML guidance term weighted by $c_a$ for enhanced predictive capability (\Cref{fig:figure_1}a) (see Supplementary Information Section 2.3). For clarity in the main text, we refer to these three weighting parameters as $c_\text{local}$, $c_\text{social}$, and $c_\text{ml}$, respectively. This approach provides an effective optimisation framework grounded in physical understanding, where particle movements follow intuitive vector mechanics by combining directional ``forces'' towards personal best positions, collective knowledge, and ML-predicted regions. Through $\alpha$-PSO's components, chemists can both understand and customise particle behaviour, allowing them to tune swarm dynamics to their specific objectives while facilitating diagnosis and correction (\Cref{fig:figure_1}b). This finely-tuned mechanistic control of swarm behaviour ensures the search process is in alignment with scientific goals and chemical expertise.

\section{Performance benchmarking on experimentally-derived datasets}
\label{sec:computational_benchmark}

\begin{figure}[H]
    \centering
    \includegraphics[width=1\linewidth]{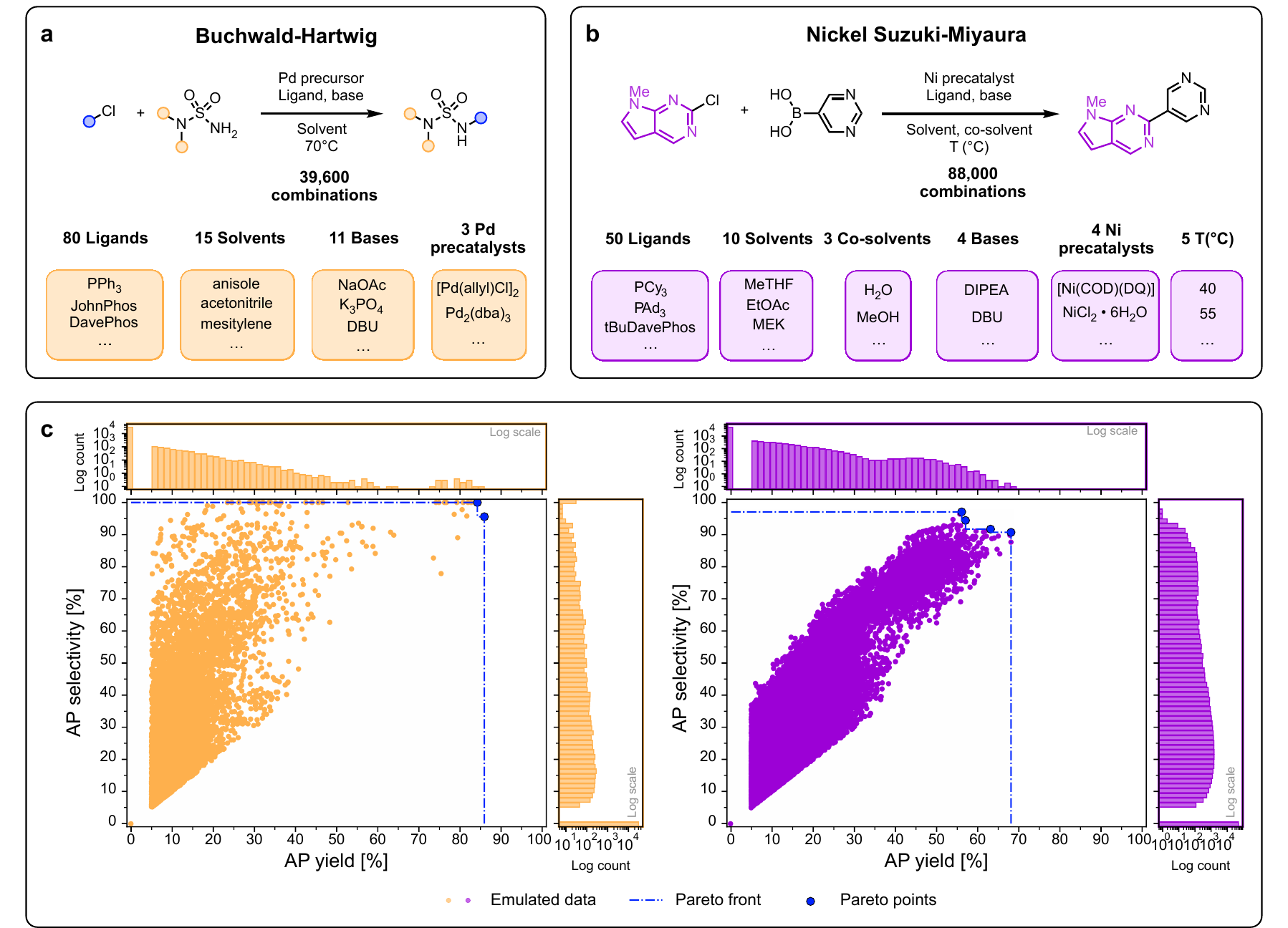}
    \caption{
    \textbf{Benchmark datasets used to assess the performance of optimisation algorithms.} \textbf{a,} Reaction condition space for the Pd-catalysed Buchwald-Hartwig benchmark dataset generated with experimental data from Sin et al.~\cite{sinHighlyParallelOptimisation2025} (see Supplementary Information Section 3.1) \textbf{b,} Reaction condition space for the Ni-catalysed Suzuki coupling benchmark dataset generated with experimental data from Sin et al.~\cite{sinHighlyParallelOptimisation2025} (see Supplementary Information Section 3.1) \textbf{c,} Distribution of the reaction objectives, AP yield [\%] and selectivity [\%] for both the Pd-catalysed Buchwald-Hartwig (left panel, orange) and Ni-catalysed Suzuki (right panel, purple) benchmark datasets. For each plot, the blue circles indicate the Pareto-optimal points representing the best AP yield-selectivity trade-offs achievable in each reaction space, with the Pareto front connecting these points (see Supplementary Information Section 1.1). Marginal histograms for each benchmark dataset display the distribution density of AP yield and selectivity values in the dataset on logarithmic scales.
    }
    \label{fig:benchmark_presentation}
\end{figure}

We first evaluated $\alpha$-PSO using two pharmaceutically relevant reaction datasets covering a broad reaction space: a Ni-catalysed Suzuki reaction and a Pd-catalysed Buchwald-Hartwig sulfonamide coupling of an active pharmaceutical ingredient (API) (\Cref{fig:benchmark_presentation}). These cross-coupling transformations, representing essential and frequently employed methodologies in modern drug synthesis~\cite{raghavanIncorporatingSyntheticAccessibility2024,fitznerWhatCanReaction2020}, were derived from HTE experimental data collected in our lab (see Supplementary Information Section 3.1)~\cite{sinHighlyParallelOptimisation2025}. Their distinct chemical landscapes provide characteristic optimisation challenges for evaluating $\alpha$-PSO's real-world performance. We hypothesised that $\alpha$-PSO could effectively leverage the large batch sizes inherent to HTE workflows to enable effective swarm dynamics and particle communication. Using these benchmark datasets, we simulated multi-objective optimisation campaigns where optimisation algorithms iteratively searched for optimal reaction conditions, evaluating their ability to identify the highest-performing combinations of area percent (AP) yield and selectivity from the known reaction condition results. All optimisation campaigns were initialised using quasi-random Sobol sampling~\cite{burhenneSamplingBasedSobol2022} (see Supplementary Information Section 2.2) to ensure well-distributed starting points across the reaction condition space for initialisation of the $\alpha$-PSO swarm. We evaluated $\alpha$-PSO against three classes of optimisation methods: (1) Bayesian optimisation methods using the Minerva~\cite{sinHighlyParallelOptimisation2025} workflow with the q-Noisy Expected Hypervolume Improvement~\cite{daultonParallelBayesianOptimization2021} (qNEHVI) acquisition function, which has demonstrated superior performance in Bayesian reaction optimisation over other approaches across multiple theoretical and empirical studies~\cite{daultonParallelBayesianOptimization2021, zhangMultiobjectiveBayesianOptimisation2024,sinHighlyParallelOptimisation2025}; (2) the canonical PSO algorithm~\cite{kennedyParticleSwarmOptimization1995} (see Supplementary Information Section 2.2); and (3) quasi-random Sobol sampling as a competitive baseline superior to pure random search~\cite{torresMultiObjectiveActiveLearning2022}. To quantitatively assess algorithm performance, we used the hypervolume indicator~\cite{guerreiroHypervolumeIndicatorProblems2022a}, which quantifies both the quality and diversity of multi-objective solutions identified (see Supplementary Information Section 1.1). We repeated optimisation experiments across 20 random seeds. All optimisation methods were run for 6 total iterations, including initialisation, with batch sizes of 24, 48, and 96, reflecting plate sizes commonly observed in HTE systems utilising solid dosing.

\begin{figure}[H]
    \centering
    \includegraphics[width=1\linewidth]{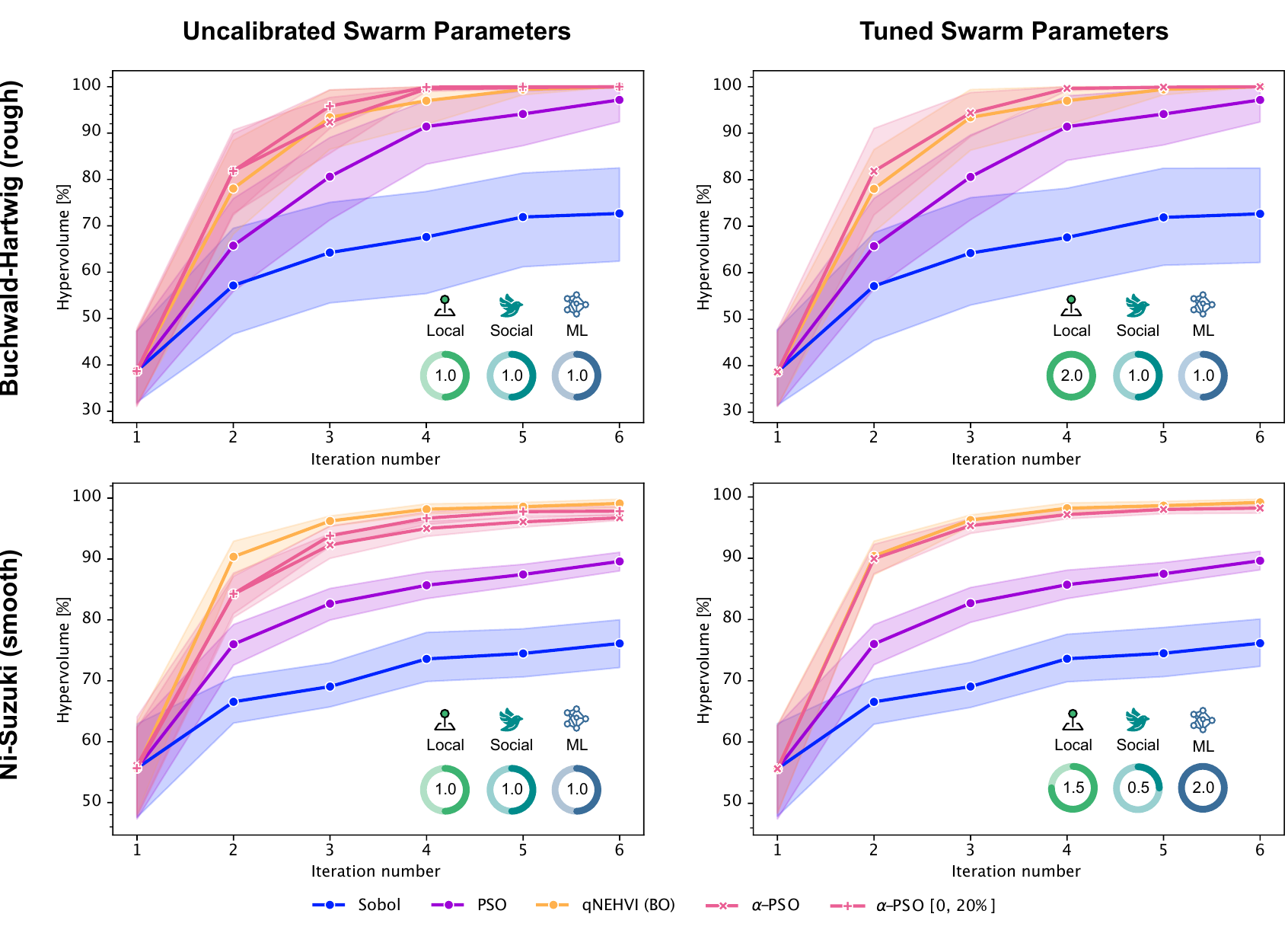}
    \caption{\textbf{Comparative performance of optimisation algorithms on Buchwald-Hartwig (top row, rough landscape) and Ni-Suzuki (bottom row, smooth landscape) benchmark datasets at batch size $q=96$.} Left panels: Default parameter configuration ($c_\text{local}=c_\text{social}=c_\text{ml}=1.0$) showing baseline performance across algorithms. Right panels: Optimised parameter configurations tailored to each landscape topology, with parameter values indicated by circular icons. Performance measured by hypervolume indicator across 20 random seeds with 95\% confidence intervals. $\alpha$-PSO [0, 20\%] indicates $\alpha$-PSO with relocation of the 20\% worst particles in the swarm to ML predicted positions (see Supplementary Information Section 2.3). Landscape-specific parameter tuning enables $\alpha$-PSO to outperform state-of-the-art Bayesian optimisation approaches (qNEHVI) on rough landscapes and achieve competitive performance on smooth landscapes.}
    \label{fig:benchmark_results}
\end{figure}

Using the Buchwald-Hartwig benchmark, we first sought to evaluate $\alpha$-PSO using simple, equally weighted parameters of $c_\text{local}=c_\text{social}=c_\text{ml}=1.0$ to characterise baseline $\alpha$-PSO performance prior to detailed parameter studies. As expected, both $\alpha$-PSO and Bayesian optimisation significantly outperformed the quasi-random Sobol baseline. Further, $\alpha$-PSO substantially outperformed the standard PSO algorithm, demonstrating the benefits of our ML augmentation term. Notably, this uncalibrated implementation of $\alpha$-PSO demonstrated superior performance compared to Bayesian optimisation in the second iteration across all batch sizes. $\alpha$-PSO demonstrated stronger overall performance at larger batch sizes ($q=96$), outperforming Bayesian optimisation (\Cref{fig:benchmark_results}). This performance advantage diminished at smaller batch sizes, achieving parity with Bayesian optimisation at $q=48$, and Bayesian optimisation showing better performance at $q=24$ (see Supplementary Information Section 4). This batch size dependence aligns with fundamental swarm intelligence principles, where larger populations enable more effective collective behaviour through enhanced information sharing across the particle swarm. Extending our investigation to the Ni-catalysed Suzuki coupling dataset revealed markedly different algorithm performance patterns when compared to the Buchwald-Hartwig dataset, though both $\alpha$-PSO and Bayesian optimisation continued to outperform the standard PSO and Sobol baselines. Using the same uncalibrated baseline parameters, $\alpha$-PSO maintained its batch size performance dependency, though Bayesian optimisation exhibited better performance overall (see Supplementary Information Section 4). These dataset-dependent optimisation outcomes prompted us to thoroughly examine the reaction landscape topology of our benchmark datasets and the influence of $\alpha$-PSO parameters on optimisation performance beyond uncalibrated configurations.

\begin{figure}[H]
    \centering
    \includegraphics[width=1\linewidth]{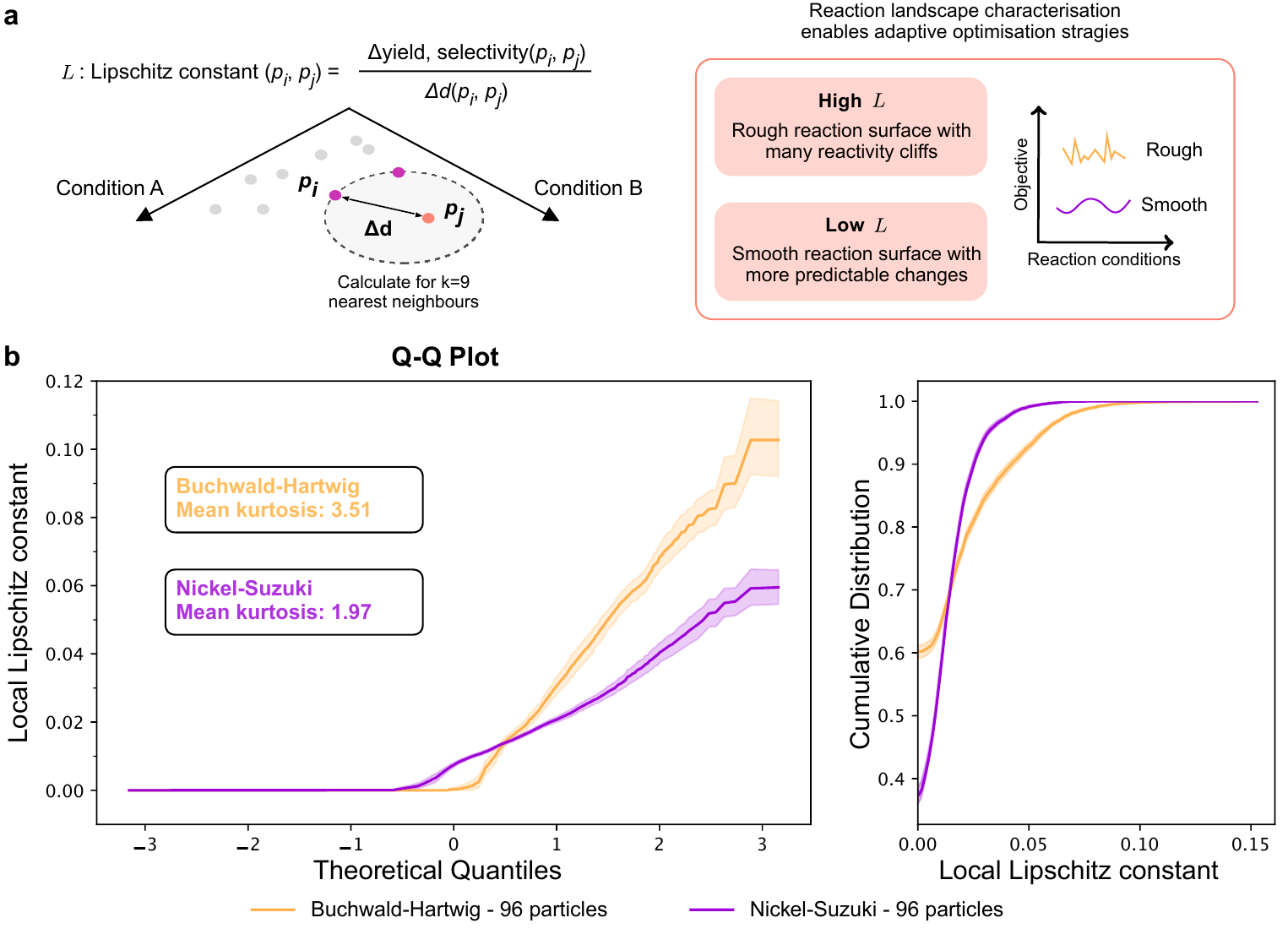}
    \caption{\textbf{Reaction landscape characterisation of the Ni-catalysed Suzuki Coupling and Pd-catalysed Buchwald-Hartwig benchmark datasets.} \textbf{a,} Conceptual framework for local Lipschitz constant analysis. Local Lipschitz constants quantify reaction landscape ``roughness'' of a dataset by measuring differences in reaction outcomes (e.g., yield and selectivity) ($\Delta y$) relative to differences in reaction conditions ($\Delta d$) for each reaction condition's $k$-nearest neighbours. $k=9$ was selected to ensure sufficient capture of local neighbourhood trends for each particle  (see Supplementary Information Section 3.2). High Lipschitz values indicate rough landscapes with reactivity cliffs where small condition changes produce large yield variations, while low values indicate smooth, predictable response surfaces. \textbf{b,} Statistical comparison of normalised local Lipschitz constant distributions using Sobol subsamples (96 reactions) drawn from each dataset to quantify characteristics of the complete reaction spaces, repeated across 20 random seeds. This enables characterisation of complete reaction condition search spaces with only a small set of experimentally evaluated points. Confidence bands for both plots represent variability across different subsampled realisations at each random seed. (Left) Q-Q plots show the distribution of Lipschitz constants with respect to theoretical quantiles from a standard normal distribution, where negative values represent lower percentiles and positive values represent higher percentiles. Statistical metrics show that the Buchwald-Hartwig dataset exhibits heavier-tailed distributions with higher kurtosis and higher maximum Lipschitz values, representing a more ``rough'' landscape topology with reactivity cliffs compared to the Ni-Suzuki coupling dataset. (Right) Cumulative distribution function plots showing the probability of encountering Lipschitz constants below given thresholds.}
    \label{fig:lipschitz_analysis_main}
\end{figure}

To quantify landscape characteristics for each dataset, we used local Lipschitz constant~\cite{jordanExactlyComputingLocal2020} analysis across $k=9$ nearest neighbours~\cite{taunkBriefReviewNearest2019}, providing a mathematical framework for measuring reaction space ``roughness'', the degree to which small changes in reaction conditions produce large variations in reaction outcomes (\Cref{fig:lipschitz_analysis_main}a). We used Sobol subsamples drawn from each dataset to quantify characteristics of the complete reaction spaces, revealing distinct landscape topologies between our benchmark datasets (\Cref{fig:lipschitz_analysis_main}b).
The Buchwald-Hartwig dataset exhibited substantially higher normalised local Lipschitz constants and kurtosis values (a measure of the prevalence of extreme values), indicating frequent reactivity cliffs, while the Ni-Suzuki dataset showed significantly lower values, reflecting smoother response surfaces. Further analysis with the complete reaction spaces confirmed that these Sobol subsamples accurately captured landscape characteristics of the full datasets, demonstrating reliable characterisation of complete datasets from a small subset of initial experiments (see Supplementary Information Section 3.2). The quantitative difference in landscape topology across datasets explains the variations in observed algorithm performance and guides landscape-dependent parameter selection.

Systematic $\alpha$-PSO parameter screening at batch size $q=96$ revealed different performance patterns for each landscape topology, with both datasets showing performance degradation when global search emphasis ($c_\text{social}$) increased excessively. On the rough Buchwald landscape (\Cref{fig:benchmark_results}), $\alpha$-PSO achieved superior performance over Bayesian optimisation across multiple parameter configurations, with $c_\text{social}$ values of $0.5$ and $1.0$ providing effective exploration-exploitation balance and $c_\text{local}$ values showing minimal impact on performance. Notably, increasing ML guidance ($c_\text{ml}$) beyond the default configuration ($c_\text{ml}=1.0$) did not improve performance and frequently degraded results, suggesting that inherent swarm dynamics sufficiently handle the navigation of rough landscapes better than ML methods. The smooth Ni-Suzuki reaction landscape displayed markedly different parameter dependencies: ML guidance at $c_\text{ml}=2.0$ appreciably enhanced performance over $c_\text{ml}=1.0$, and $\alpha$-PSO showed heightened sensitivity to global search parameters, with performance degrading when $c_\text{social}$ was increased past 0.5 regardless of other parameters. This enhanced benefit from ML guidance on smooth landscapes aligns with expectations, as the more predictable and smoothly varying response surfaces enable ML models to provide reliable directional guidance, complementing swarm dynamics. Additionally, both benchmarks benefited from $\alpha$-PSO particle reinitialisation, with the relocation of approximately 20\% of the worst-performing particles optimising the balance between performance improvement and maintaining swarm coherence (\Cref{fig:benchmark_results}) (see Supplementary Information Section 2.3). Optimal parameter choice on both datasets enabled $\alpha$-PSO to outperform Bayesian optimisation on the Buchwald-Hartwig benchmark and achieve parity with Bayesian optimisation on the Ni-Suzuki benchmark under multiple optimised configurations (\Cref{fig:benchmark_results}).

\section{Prospective application: Sulfonamide coupling}

\begin{figure}[H]
    \centering
    \includegraphics[width=1\linewidth]{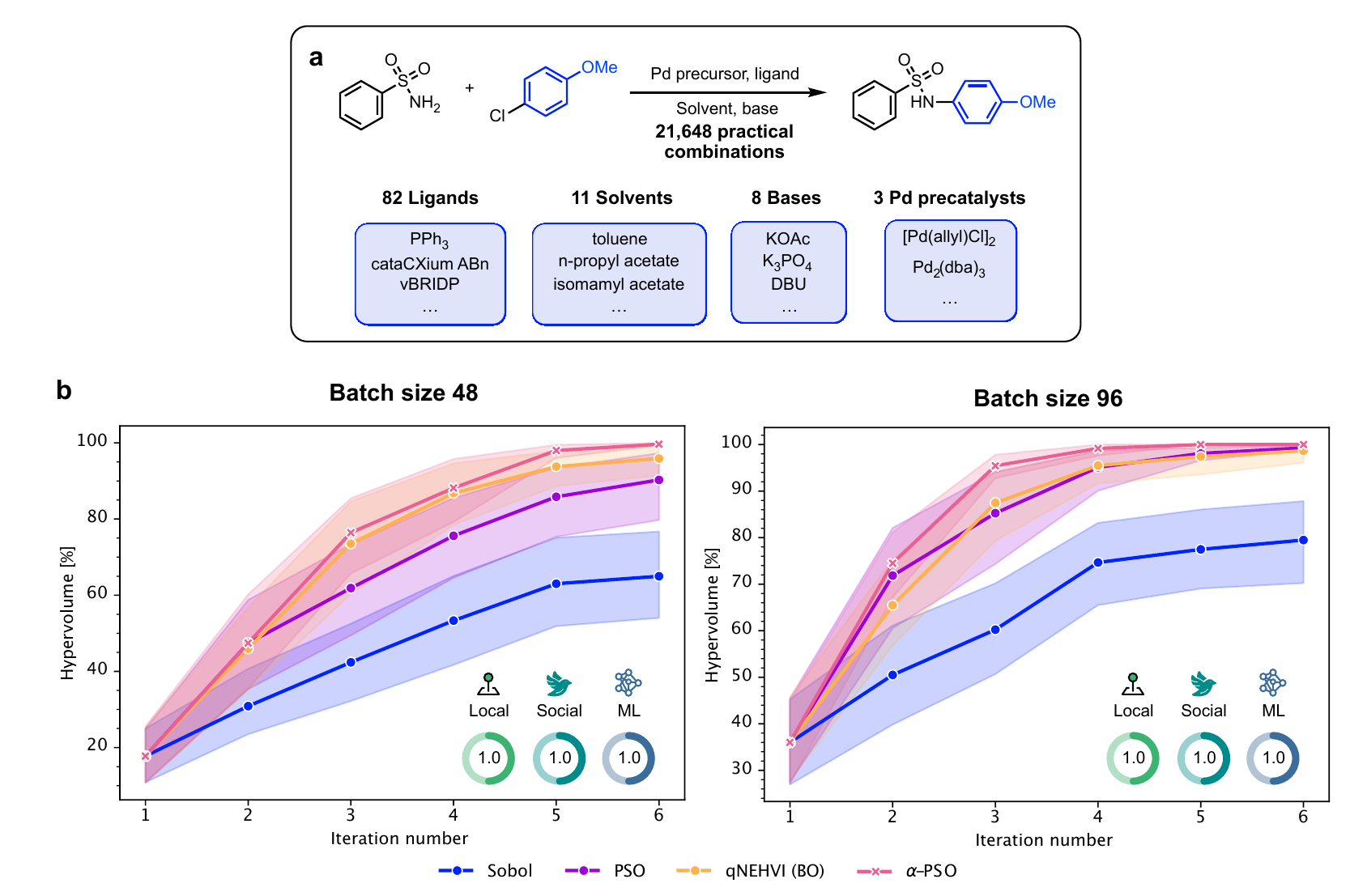}
    \caption{\textbf{Prospective validation of $\alpha$-PSO on a Pd-catalysed sulfonamide coupling reaction.} \textbf{a,} Reaction condition space for the Pd-catalysed sulfonamide coupling comprising 82 ligands, 11 solvents, 8 bases, and 3 palladium precursors (21,648 total combinations). Examples in each reaction component class are shown. \textbf{b,} Comparative performance of optimisation algorithms across batch sizes $q=48$ (left) and $q=96$ (right) particles, quantified by the hypervolume indicator for multi-objective optimisation of AP yield and selectivity. $\alpha$-PSO was configured with default baseline parameters ($c_\text{local}=c_\text{social}=c_\text{ml}=1.0$), based on landscape roughness analysis showing extreme roughness of this reaction. Performance plots show mean values across 20 random seeds with the shaded area representing 95\% confidence intervals. $\alpha$-PSO demonstrates superior performance compared to Bayesian optimisation (qNEHVI) at both batch sizes, validating the landscape-guided parameter selection framework and confirming the effectiveness of default parameters on rough reaction landscapes. Canonical PSO without any ML components achieved comparable performance to Bayesian optimisation at batch size $q=96$, demonstrating that simple swarm dynamics can effectively navigate rough landscapes.}
    \label{fig:sulfonamide_fig}
\end{figure}

With insights into $\alpha$-PSO's landscape-dependent behaviour, we sought to further validate both our characterisation of reaction topology patterns and understand $\alpha$-PSO parameter behaviour on new reaction systems. We selected an industrially relevant Pd-catalysed sulfonamide coupling reaction as a prospective case study, predicting that the hard coupling partners would exhibit similarly rough landscape characteristics to our Buchwald-Hartwig benchmark dataset in \Cref{sec:computational_benchmark}. We defined a reaction search space comprising 82 ligands, 11 solvents, 8 bases, and 3 palladium precursors (a total of 21,648 reaction condition combinations), towards the multi-objective optimisation of AP yield and selectivity (\Cref{fig:sulfonamide_fig}a). Given the strong performance of $\alpha$-PSO with default, untuned parameters on rough datasets ($c_\text{local}=c_\text{social}=c_\text{ml}=1.0$), we employed these same baseline parameters to test their effectiveness on this new reaction system. Within this reaction search space, we collected an experimental dataset of 585 reaction conditions using HTE, generating a virtual benchmark for algorithm comparison between $\alpha$-PSO and BO to enable statistically significant evaluation across multiple random initialisations and seeds (see Supplementary Information Section 7).

Quantitative landscape analysis of an experimentally subsampled Sobol batch of 96 reactions from the search space with local Lipschitz constants showed that the experimental sulfonamide coupling dataset exhibited rough landscape characteristics with elevated kurtosis~\cite{hoDescriptiveStatisticsModern2015} (13.29) values, indicating heavy-tailed distributions with frequent outliers that closely aligned with our computational Buchwald-Hartwig benchmark (Supplementary Figure 37). Consistent with our assumptions based on prior benchmarking on rough datasets, $\alpha$-PSO with default parameters demonstrated superior performance over Bayesian optimisation at both batch sizes $q=48$ and $q=96$ (\Cref{fig:sulfonamide_fig}b). Remarkably, even canonical PSO without any ML components achieved comparable performance to Bayesian optimisation at batch size $q=96$, demonstrating that simple swarm dynamics can effectively navigate rough landscapes. These results demonstrate that our landscape topology analysis framework can guide effective parameter selection, enabling $\alpha$-PSO to outperform state-of-the-art Bayesian optimisation methods.

\section{Prospective application: Suzuki-Miyaura reaction}

\begin{figure}[H]
    \centering
    \includegraphics[width=1\linewidth]{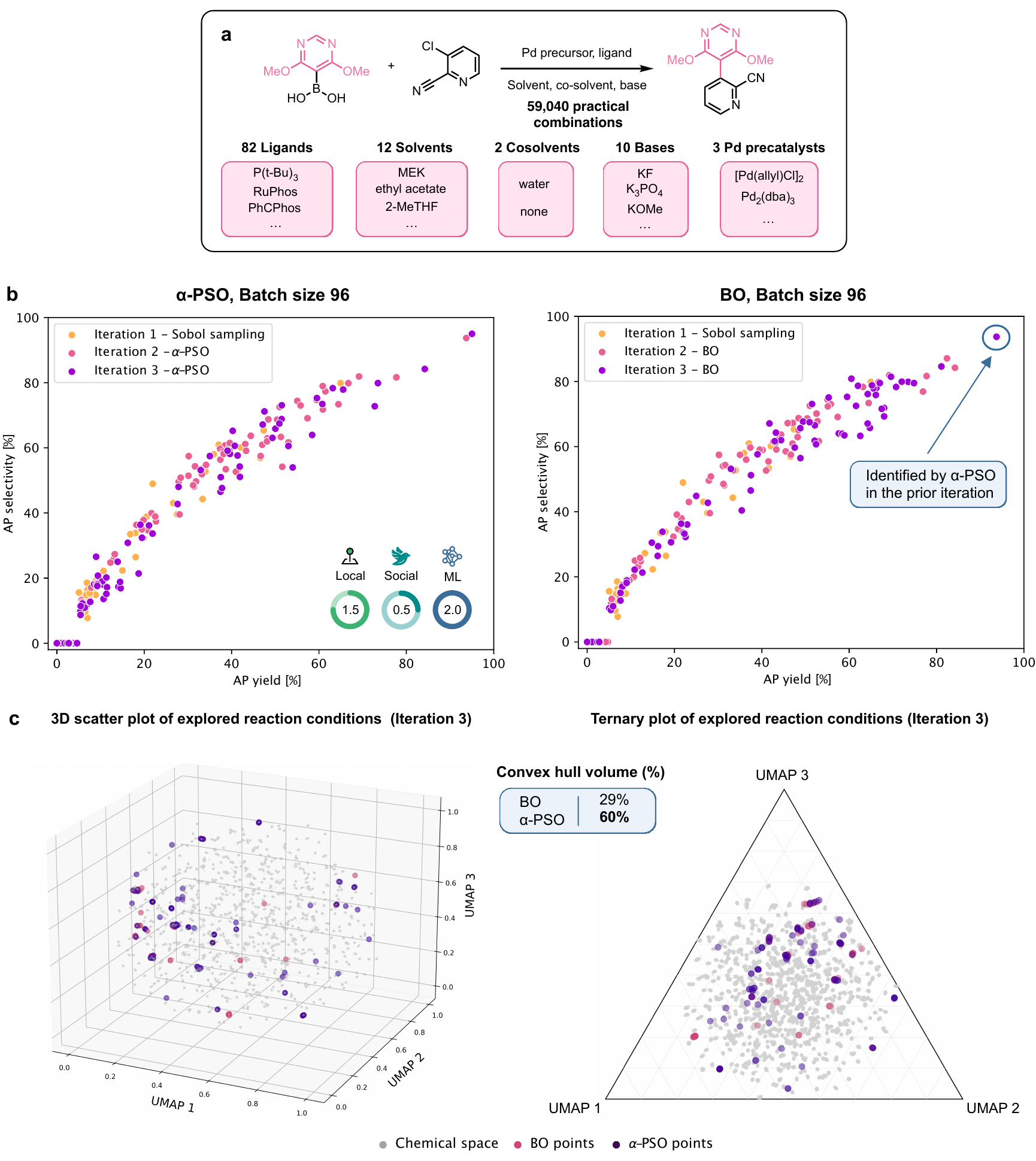}
    \caption{\textbf{Prospective validation of $\alpha$-PSO on a Pd-catalysed Suzuki coupling reaction. a,} Reaction condition space for the Pd-catalysed Suzuki coupling between electron-deficient pyrimidine and pyridine heterocycles, comprising 82 monophosphine ligands, 12 solvents, 2 co-solvents, 10 bases, and 3 palladium precursors (59,040 total combinations). Representative examples of each reaction component class are shown. \textbf{b,} Direct head-to-head experimental comparison showing reaction outcomes across three iterations. $\alpha$-PSO (left) was configured with smooth landscape parameters ($c_\text{local}=1.5$, $c_\text{social}=0.5$, $c_\text{ml}=2.0$) based on initial landscape analysis, while Bayesian optimisation (right) used qNEHVI acquisition function. $\alpha$-PSO consistently identified batch-best conditions at each iteration, achieving 94\% AP yield and selectivity in iteration 2 versus 84\% for Bayesian optimisation, and 95\% versus 94\% in iteration 3. Notably, optimal conditions found by Bayesian optimisation in iteration 3 had already been discovered by $\alpha$-PSO in iteration 2. \textbf{c,} 3D scatter plot (left) and ternary plot (right) of reaction conditions explored by each algorithm in iteration 3, projected onto three-dimensional UMAP~\cite{mcinnesUMAPUniformManifold2020} chemical space. Both visualisations show that $\alpha$-PSO exhibits notably more dispersed exploration strategies, sampling broader and more diverse regions of the reaction space while maintaining effective convergence on high-performing areas. In contrast, Bayesian optimisation concentrates suggestions within predicted high-performance regions, demonstrating the direct connection between $\alpha$-PSO parameter choices (enhanced local exploration with increased $c_\text{local}$) and observable optimisation behaviour. Convex hull analysis confirms this observation, showing that $\alpha$-PSO explored 60\% of the chemical space volume compared to 29\% for Bayesian optimisation.}
    \label{fig:case_study_2}
\end{figure}

Having shown $\alpha$-PSO's effectiveness through benchmarking studies, we next conducted a direct head-to-head experimental comparison between $\alpha$-PSO and Bayesian optimisation. We selected a challenging Pd-catalysed Suzuki coupling reaction with electron-deficient and sterically hindered substrates as our test system. This transformation represents a pharmaceutically relevant coupling between pyrimidine and pyridine heterocycles, ubiquitous scaffolds in drug discovery~\cite{vitakuAnalysisStructuralDiversity2014,heraviPrescribedDrugsContaining2020}, where the electron-deficient nature of both coupling partners and potential nitrile hydrolysis under reaction conditions present significant synthetic challenges.
We defined a search space comprising 82 monophosphine ligands, 12 solvents, 10 bases, 3 palladium precursors, and 2 co-solvents, with 59,040 unique reaction conditions to explore this chemical transformation (\Cref{fig:case_study_2}a).

Following initialisation of the first HTE batch (96 experiments) with Sobol sampling, we applied our landscape analysis framework to this initial experimental data to guide $\alpha$-PSO parameter selection. Landscape topology analysis showed smooth characteristics consistent with our Ni-catalysed Suzuki coupling benchmark dataset: the experimental data showed moderately low kurtosis (1.93) values with maximum Lipschitz constants reaching similar ranges ($\sim$0.07 compared to $\sim$0.06 for the Ni-catalysed virtual benchmark), confirming the predominance of low-Lipschitz regions where reaction conditions produce proportional and predictable changes in outcomes (Supplementary Figure 41). Based on these landscape characteristics, we configured $\alpha$-PSO with the optimal parameters identified in \Cref{sec:computational_benchmark} for smooth topologies: increased personal best emphasis ($c_\text{local} = 1.5$), reduced global search ($c_\text{social} = 0.5$), and enhanced ML guidance ($c_\text{ml} = 2.0$), directly comparing with Bayesian optimisation using the qNEHVI~\cite{daultonParallelBayesianOptimization2021} multi-objective acquisition function in the Minerva~\cite{sinHighlyParallelOptimisation2025} workflow. Direct experimental comparison at each iteration displayed distinct algorithmic behaviours. Most critically for practical reaction optimisation, $\alpha$-PSO consistently identified the batch-best reaction conditions at each experimental iteration (\Cref{fig:case_study_2}b). In iteration 2, $\alpha$-PSO discovered conditions achieving 94\% area percent yield and selectivity compared to Bayesian optimisation's best of 84\%, while iteration 3 yielded $\alpha$-PSO conditions with 95\% area percent yield versus 94\% for Bayesian optimisation (\Cref{fig:case_study_2}b). Notably, the optimal conditions identified by Bayesian optimisation in iteration 3 had already been discovered by $\alpha$-PSO in the previous iteration, demonstrating $\alpha$-PSO's ability to identify the top-performing reaction conditions more rapidly. While these results represent a single head-to-head comparison, they demonstrate the practical effectiveness of $\alpha$-PSO in real-world experimental settings, consistent with the performance observed in our computational benchmarks.

Retrospective analysis uncovered distinct differences in optimisation behaviour between the algorithms. 3D scatter plots and ternary visualisation plots of the explored reaction conditions by each algorithm, particularly in the third iteration, showed that $\alpha$-PSO employed more dispersed search strategies, exploring broader and diverse regions of the chemical space (\Cref{fig:case_study_2}c). Quantitative convex hull analysis confirmed this broader exploration, with $\alpha$-PSO covering 60\% of the accessible chemical space volume compared to only 29\% for Bayesian optimisation. This enhanced local exploration, resulting from increased local particle best emphasis ($c_\text{local}=1.5$) and reduced global search parameters ($c_\text{social}=0.5$), directly translates $\alpha$-PSO parameter choices to tangible, observable optimisation behaviour, establishing the connection between $\alpha$-PSO configuration and experimental outcomes. While Bayesian optimisation concentrated suggestions within predicted high-performance regions, $\alpha$-PSO investigated more diverse and potentially low-yielding regions across the reaction space while maintaining effective convergence on the most high-performing areas. This demonstrates how landscape-informed $\alpha$-PSO tuning can yield a practically advantageous search strategy, enabling both rapid discovery of optimal conditions and a more thorough exploration of the chemical space.

\section{Discussion}

$\alpha$-PSO provides an effective approach for parallel chemical reaction optimisation where simple physics-based rules augmented with ML provide an intuitive and performant method. Evaluation of $\alpha$-PSO across two pharmaceutically relevant reaction benchmarks demonstrates competitive performance with state-of-the-art Bayesian optimisation methods, while two prospective HTE optimisation campaigns showed $\alpha$-PSO identifying optimal reaction conditions more rapidly, reaching 94 AP yield and selectivity within two iterations for a challenging heterocyclic Suzuki reaction and achieving superior performance in a Pd-catalysed sulfonamide coupling. Notably, $\alpha$-PSO delivers strong performance while maintaining straightforward swarm dynamics connected to experimental observables, rather than relying purely on learned probabilistic abstractions. Particle movements result from directional forces towards clearly specified reaction conditions in the search space, emerging from explicit combinations of personal experience ($v_1$), collective knowledge ($v_2$), and ML predictive guidance ($v_a$) with tunable coefficients $c_\text{local}$, $c_\text{social}$, and $c_\text{ml}$. Our theoretical framework using local Lipschitz constants enables landscape-specific optimisation strategies, where rough landscapes characterised by reactivity cliffs benefit from inherent swarm dynamics that handle discontinuous response surfaces, while smooth landscapes leverage more ML guidance to exploit smoothly varying and predictable trends.

In this work, we focused exclusively on industrially applicable cross-coupling reactions, representing a limited subset of synthetic transformations. Systematic validation across diverse reaction classes would further establish the generalisability of our approach. However, $\alpha$-PSO's gradient-free nature and lack of assumptions about functional form suggest broader applicability to general multi-objective optimisation problems. Future developments could explore dynamic search spaces where the space of reaction conditions evolves during campaigns. Additionally, adaptive $\alpha$-PSO parameter strategies could modify swarm behaviour across each iteration of optimisation rather than maintaining fixed parameters throughout the entire campaign. $\alpha$-PSO is made available as an open-source repository alongside the complete HTE datasets ($989$ reactions) from both prospective experimental campaigns in the Simple User-Friendly Reaction Format (SURF)~\cite{nippaSimpleUserFriendlyReaction2024}.

\section*{Methods}\label{sec:methods}

Complete descriptions of the canonical PSO algorithm, $\alpha$-PSO algorithm, all algorithmic approaches used in this study, emulated benchmarking datasets, and HTE experimental procedures are detailed in the Supplementary Information.

\section*{Data availability}

All reaction datasets evaluated in benchmarking studies, reaction condition spaces, and experimentally collected datasets are included in the manuscript, Supplementary Information, and the accompanying public GitHub repository.

\section*{Code availability}
The custom code used in this study is made available in a public GitHub repository under the MIT open-source licence: \href{https://github.com/schwallergroup/alphaswarm}{github.com/schwallergroup/alphaswarm}.

\section*{Acknowledgements}

We thank T. V\"{o}gtlin, M. M\"{u}ller, T. Chi, E. Gianinazzi, and D. Huck from the High-Output Reaction Screening System (HORSS) team at F. Hoffmann-La Roche Ltd. for experimental and analytical support. We thank the Global Internship Program in Innovation \& Sustainability (IP2TIS) 2024 (\href{https://go.roche.com/ip2tis\_24}{go.roche.com/ip2tis\_24}) for financial support of this project. J.W.S., R.P.B., K.P., and R.B. thank Roche and its Technology Innovation and Science (TIS) initiative for financial support. P.S. acknowledges support from NCCR Catalysis (grant no. 225147), a National Centre of Competence in Research funded by the Swiss National Science Foundation.

\section*{Author contributions}
R.S. and J.W.S. contributed to the conceptualisation, methodology, code development, analysis, visualisation, and writing of the manuscript. R.P.B. and K.P. contributed to the supervision and writing of the manuscript. R.B. contributed to conceptualisation, supervision, experimental results, and writing of the manuscript. P.S. contributed to conceptualisation, methodology, supervision, and writing of the manuscript. 

\section*{Competing interests}
R.S., J.W.S., R.P.B, K.P., and R.B. declare potential financial and non-financial conflicts of interest as employees of F. Hoffmann-La Roche Ltd. The other authors declare no competing interests.

\pagebreak

\printbibliography

\includepdf[pages=-]{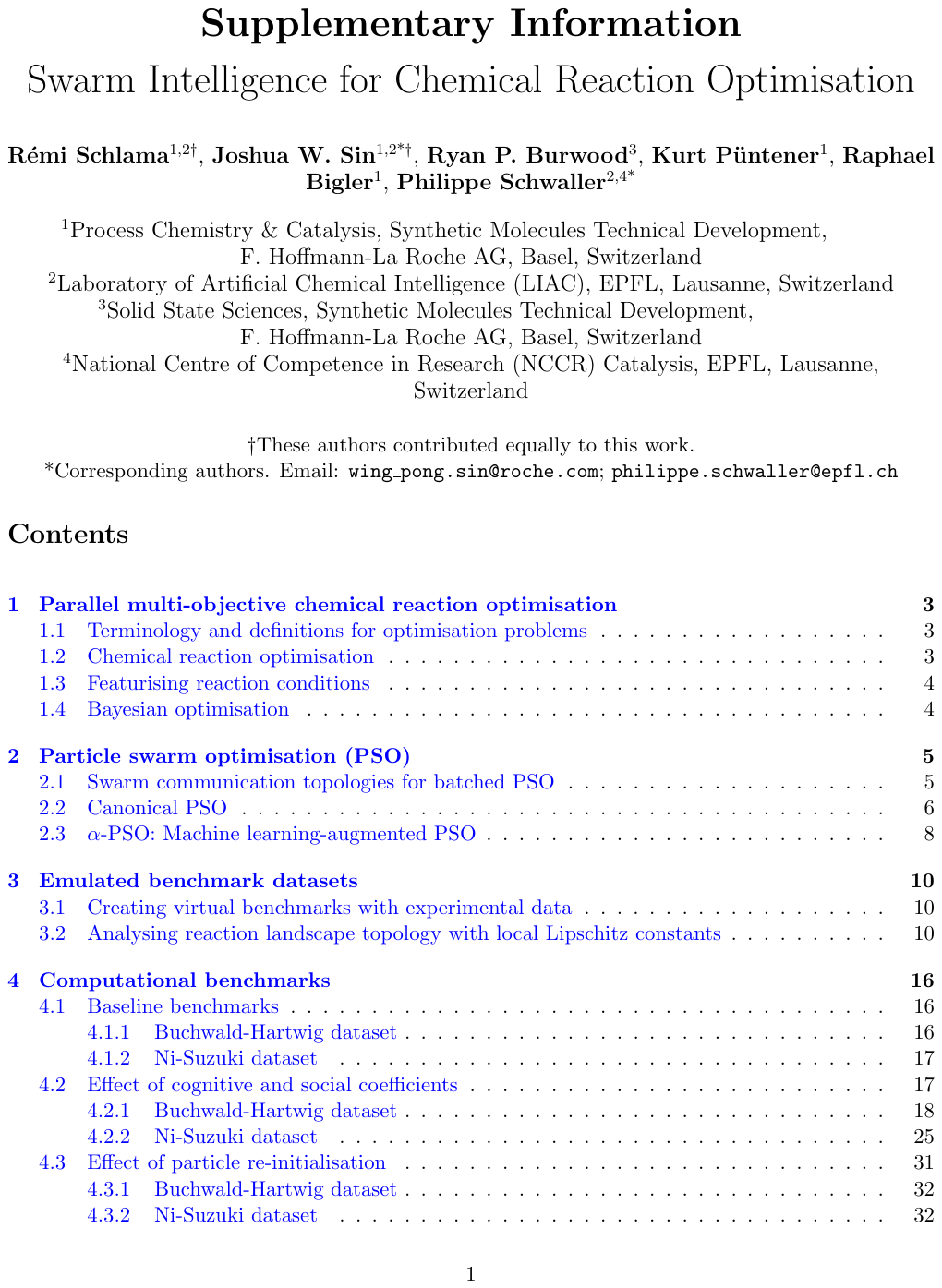}

\end{document}